\documentclass[aps,10pt,twocolumn,showpacs,amsmath,amssymb,prl,superscriptaddress]{revtex4-1}

\usepackage{graphicx}
\usepackage{amssymb}

\def\ba{\begin{array}}
\def\ea{\end{array}}
\def\beqn{\begin{eqnarray}}
\def\l{\left}
\def\rr{\right}
\def\be{\begin{equation}}
\def\ee{\end{equation}}
\def\H{\mathcal{H}}

\def\eeqn{\end{eqnarray}}

\begin{document}

\title{Adiabatic quantum motors}

\author{Ra\'ul Bustos-Mar\'un}
\affiliation{\mbox{Dahlem Center for Complex Quantum Systems and Fachbereich
Physik, Freie Universit\"at Berlin, 14195 Berlin, Germany}}

\author{Gil Refael} 
\affiliation{Department of Physics, California Institute of Technology, Pasadena, CA 91125, USA}
\affiliation{\mbox{Dahlem Center for Complex Quantum Systems and Fachbereich
Physik, Freie Universit\"at Berlin, 14195 Berlin, Germany}}

\author{Felix von Oppen}
\affiliation{\mbox{Dahlem Center for Complex Quantum Systems and Fachbereich
Physik, Freie Universit\"at Berlin, 14195 Berlin, Germany}}
\affiliation{Department of Physics, California Institute of Technology, Pasadena, CA 91125, USA}

\date{\today}
\begin{abstract}
When parameters are varied periodically, charge can be pumped through a mesoscopic conductor without applied bias. Here, we consider the inverse effect in which a transport current drives a periodic variation of an adiabatic degree of freedom. This provides a general operating principle for adiabatic quantum motors, for which we develop a comprehensive theory. We relate the work performed per cycle on the motor degree of freedom to characteristics of the underlying quantum pump and discuss the motors' efficiency. Quantum motors based on chaotic quantum dots operate solely due to quantum interference, motors based on Thouless pumps have ideal efficiency.
\end{abstract}
\pacs{}
\maketitle

{\it Introduction.---}Popular culture has long been fascinated with microscopic and nanoscopic motors. Perhaps best known is the contest announced by Richard Feynman, who promised a $\$1000$ prize to the developer of an engine that fits a cube of sides 1/64'' \cite{Feynman}. While this feat was carried out shortly thereafter, in 1960, and did not produce an intellectual breakthrough, Feynman's contest has continued to provide tremendous inspiration to the field of nanotechnology. A prototypical nanomotor was unveiled in 2003, using tiny gold leaves mounted on multi-walled carbon nanotubes, with the carbon layers themselves carrying out the motion \cite{Zettl}. The motor was driven through AC actuation, and basically relied on classical physics for their operation. As the dimensions of motors are reduced, however, it is natural to expect that quantum mechanics could be used to operate and to optimize nanomotors. In fact, cold-atom-based AC-driven quantum motors have been explored in Refs.\ \cite{Haenggi,Weitz}.

Nanomotors can also be actuated by DC driving \cite{Bailey08,Dundas09,Qi09}. A general strategy towards realizing a DC nanoscale motor is based on operating an electron pump in reverse. Consider an electron pump in which the periodic variation of parameters (such as shape, gate voltage, or tunneling strength) originates from the adiabatic motion of, say, a mechanical rotor degree of freedom. To operate this pump as a motor, an applied  bias voltage produces a charge current through the pump which, in turn, exerts a force on the mechanical rotor. The existence of quantum pumps \cite{Brouwer,Switkes} suggests that by this operating principle, quantum mechanics can be put to work in DC-driven nanomotors. Here, we develop a theory of such adiabatic quantum motors, expressing the work performed per cycle in terms of characteristics of the pump on which the motor relies and discussing the efficiency of quantum motors in general terms. 

Our theory relies on progress in the understanding of adiabatic reaction (or current-induced) forces \cite{Dundas09,Lu10,Bode11,Bode12,Thomas12} which applies when the mechanical motor degree of freedom is slow compared to electronic time scales and can be treated as classical. Conventionally, adiabatic reaction forces acting on the slow degree of freedom are considered for closed quantum systems \cite{Berry}. This has recently been extended to situations where the fast degrees of freedom constitute a quantum mechanical scattering problem and thus to mesoscopic conductors \cite{Bode11,Bode12,Thomas12}. The resulting expressions for the reaction forces in terms of the scattering matrix of the mesoscopic conductor allow one to explore the relations to quantum pumping in general terms.

\begin{figure}[t]
\includegraphics[width=4cm, keepaspectratio=true]{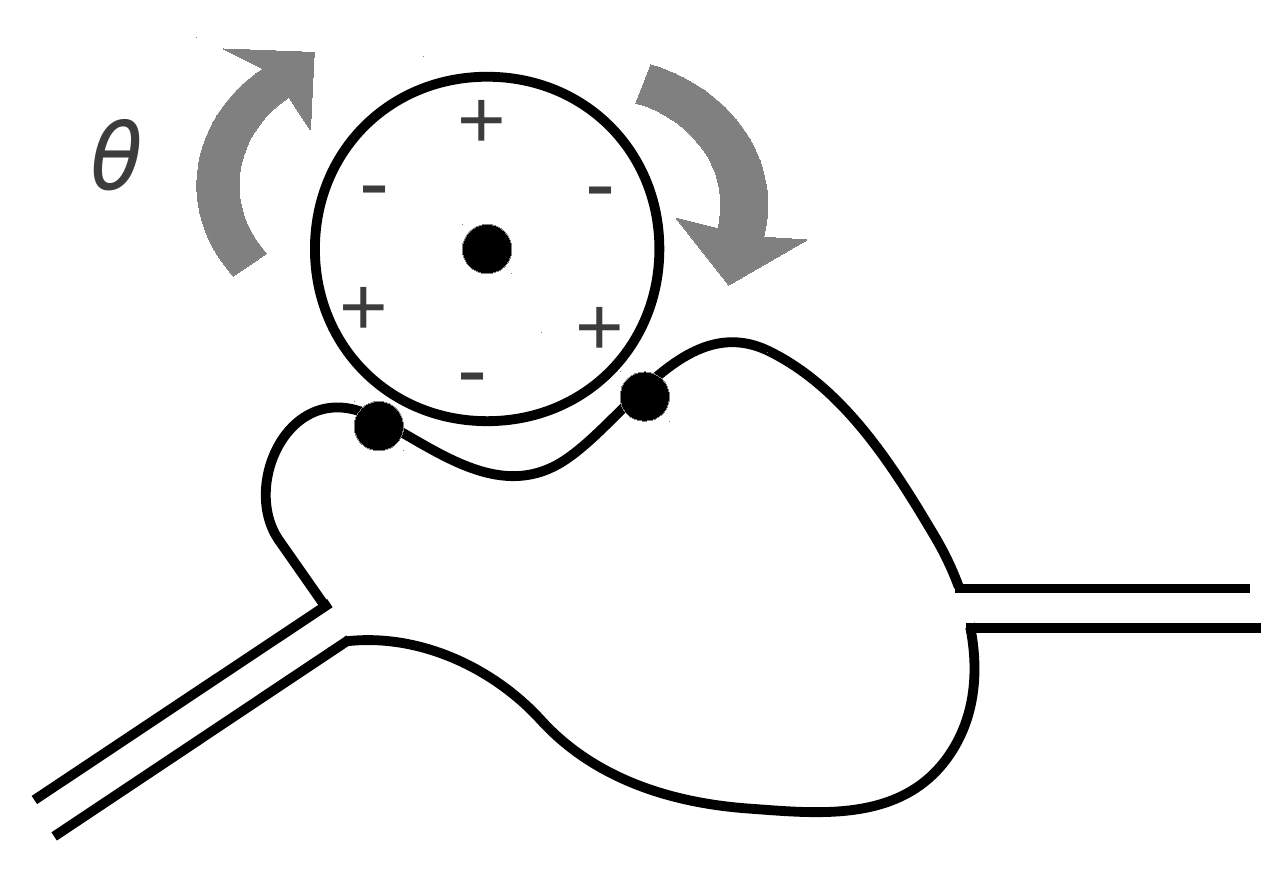}\includegraphics[width=4.5cm, keepaspectratio=true]{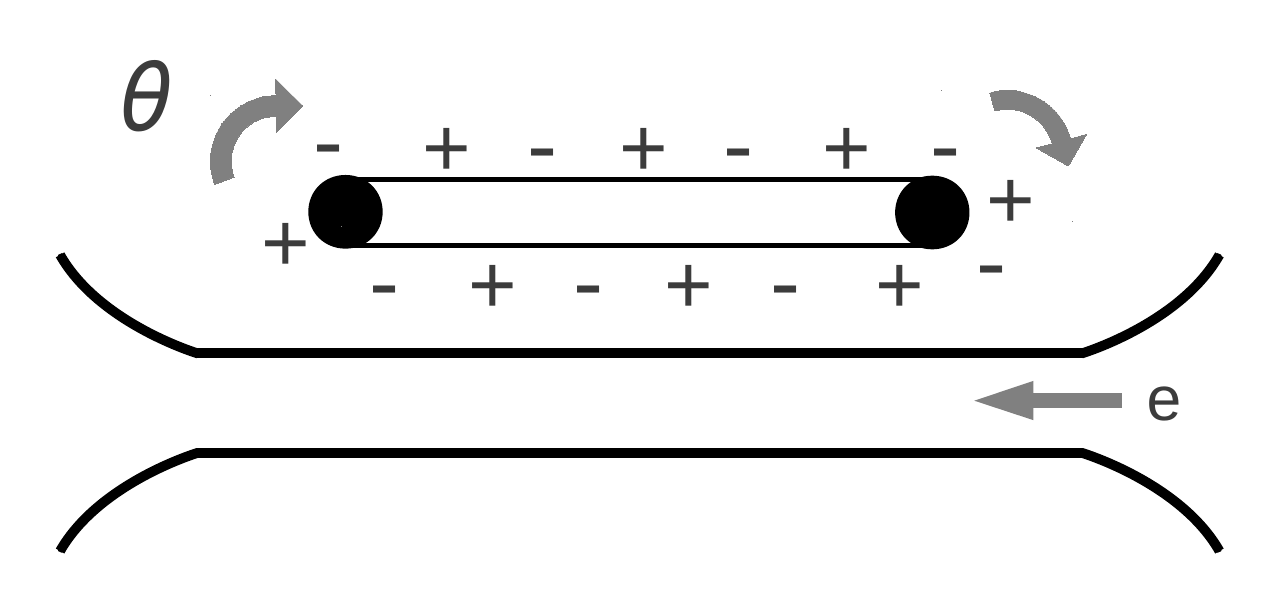}
\caption{Generic adiabatic quantum motors building on (a) a quantum pump based on a chaotic quantum dot and (b) a Thouless pump. When a voltage is applied to the pump, the current 'turns the wheel' and makes the phase angle $\theta$ wind.
\label{genericmotor}}
\end{figure}

Before developing our general theory, we sketch two conceptual examples of adiabatic quantum motors in Fig.\ \ref{genericmotor}. One motor is based on a chaotic quantum dot operated as a pump \cite{Brouwer, Switkes}, as illustrated in Fig.\ \ref{genericmotor}(a). In this motor, the time-dependent gate voltages varying the shape of the quantum dot are provided by a periodic set of charges situated around the rim of a wheel which approach and modify the quantum dot in two locations. A current flowing through the quantum dot will then produce a rotation of the wheel. Alternatively, we could base a quantum motor on a Thouless pump. A schematic of such a motor is shown in Fig.\ \ref{genericmotor}(b). A single-channel quantum wire is located next to a conveyor belt with periodic attached charges (alternatively, a cogwheel with periodically spaced and electrically charged teeth). The charges induce a periodic potential in the quantum wire which slides as the conveyor belt or cogwheel turns. It is well-known from the seminal works of Thouless \cite{Thouless} that when the Fermi energy lies in an energy gap, such pumps transport integer amounts of charge per cycle (i.e., when the periodic potential slides by one period). An alternative physical realization of a Thouless motor is based on a helical wire in an electric field \cite{Qi09}. 

{\it Output power of adiabatic quantum motors.---}We start by deriving a general expression for the output power of an adiabatic quantum motor. The motor consists of a mesoscopic conductor with left (L) and right (R) lead, described within the independent-electron approximation by an electronic scattering matrix. In the adiabatic quantum motors of Fig.\ \ref{genericmotor}, the mesoscopic conductor is coupled to a single (classical) angle degree of freedom $\theta$, as described through the dependence $S({\theta})$ of the $S$-matrix on the motor coordinate ${\theta}$. More generally, the mesoscopic conductor could be coupled to several mechanical motor degrees of freedom $X_\nu$ ($\nu=1,2\ldots N$) so that $S=S({\bf X})$. 

Retaining the dependence on several mode coordinates ${\bf X}$ for generality, the adiabatic reaction force ${\bf F}({\bf X})$ on the motor degrees of freedom can be expressed in terms of the $S$-matrix of the mesoscopic conductor \cite{Bode11,Bode12,Thomas12}, 
\begin{equation} \label{force}
 F_\nu({\bf X}) = \sum_\alpha \int \frac{d\epsilon}{2\pi i}  f_\alpha 
 {\rm Tr} \left(\Pi_\alpha S^\dagger \frac{\partial S}{\partial X_\nu} \right).
\end{equation}
Here, $f_\alpha(\epsilon)$ denotes the Fermi distribution function in lead $\alpha = {\rm L,R}$ with chemical potential $\mu_\alpha$ and $\Pi_\alpha$ is a projector onto the scattering channels in lead $\alpha$. It has been shown \cite{Dundas09,Lu10,Bode11,Bode12,Thomas12} that this adiabatic reaction force need not be conservative when the electronic conductor is out of equilibrium. Thus, the work per cycle performed by this force is nonzero and given by
\begin{equation}
  W_{\rm out} = \oint d{\bf X}\cdot {\bf F}({\bf X}) .
  \label{OutWork}
\end{equation}
Note that in the absence of an applied bias, $W_{\rm out}=0$ (i.e., the force is conservative). In this case, $f_\alpha(\epsilon)=f(\epsilon)$ and $\sum_\alpha \Pi_\alpha = {\bf 1}$, and inserting Eq.\ (\ref{force}) into Eq.\ (\ref{OutWork}) yields
\begin{equation}
W_{\rm out} = \oint d{\bf X}\cdot \nabla_{\bf X} \int \frac{d\epsilon}{2\pi i}\,  f(\epsilon)\, 
{\rm Tr} \ln S(\epsilon) = 0 .
\end{equation}
The work performed by the adiabatic quantum motor per cycle is nonzero when a finite bias $V$ is applied. In linear response, Eq.\ (\ref{force}) yields
\begin{eqnarray}
W_{\rm out} = \frac{ieV}{4\pi }\oint d{\bf X}\cdot \!\!\int {d\epsilon} f^\prime(\epsilon) 
{\rm Tr} [ (\Pi_L - \Pi_R) S^\dagger \frac{\partial S}{\partial {\bf X}} ],
\label{Eq4}
\end{eqnarray}
where we used that $W_{\rm out}=0$ in equilibrium and expanded Eq.\ (\ref{force}) to linear order in the applied bias $V$.

Using Brouwer's formula \cite{Brouwer} which expresses the charge $Q_p$ pumped during one cycle of ${\bf X}$ in terms of the electronic $S$-matrix $S({\bf X})$, the right-hand side of Eq.\ (\ref{Eq4}) can be identified as 
\begin{equation}
W_{\rm out} = Q_p V.
\label{output}
\end{equation}
Remarkably, the output of the nonequilibrium device is described by $Q_p$, which characterizes the underlying quantum pump in {\it equilibrium}. Eq.\ (\ref{output}) shows that the mechanical output of the motor per cycle originates from the fact that a charge $Q_p$ is pumped through the system with every revolution of the motor, and that this pumped charge gains an electrical energy $Q_pV$ due to the applied bias. Thus, the average output power of the motor is 
\begin{equation}
   P_{\rm out} = Q_p V/\tau,
   \label{OutputP}
\end{equation}
where $\tau$ denotes the motor's cycle period. We also emphasize that Eq.\ (\ref{output}) identifies quantum pumping as the physical origin of the nonconservative nature of the adiabatic reaction force in Eq.\ (\ref{force}).

{\it Efficiency of adiabatic quantum motors.---}The applied bias $V$ induces a slowly-varying DC charge current $I$ in the adiabatic quantum motor. Thus, on average, operation of the motor requires an input power of $P_{\rm in} = \overline{I}V$. (The overline denotes an average over a single cycle). The efficiency $\eta $ of the adiabatic quantum motor is then naturally defined as the ratio of output to input power, 
\begin{equation}
  \eta = P_{\rm out}/P_{\rm in} = Q_p/\overline{I}\tau.
\end{equation}
Here, we have used Eq.\ (\ref{OutputP}) in the second equality. 

For adiabatic motor degrees of freedom, the current $I$ is made up of two contributions: the pumped charge and the transport current induced by the applied bias $V$. If $G({\bf X})$ denotes the conductance of the device for fixed ${\bf X}$, the linear-response current averaged over one cycle is 
\begin{equation}
 \overline{I} = \overline{G({\bf X})} V + \frac{Q_p}{\tau}.
\label{current}
\end{equation}
Note that the pumping current also depends on voltage through the motor's operating frequency (as characterized by $\tau$). We note in passing that this expression can be obtained more formally, see Ref.\ \cite{Bode11}. 

With Eq.\ (\ref{current}), the quantum motor's efficiency becomes
\begin{equation}
  \eta = \frac{1}{1+ \overline{G} V\tau/Q_p }.
  \label{Efficiency}
\end{equation}
Interesting conclusions can be drawn directly from this expression: (i) Quantum motors can operate entirely on the basis of quantum interference and become ineffective due to phase-breaking processes, justifying the term {\em quantum} motor. A conceptually interesting example is the motor in Fig.\ \ref{genericmotor}(a) which is based on a chaotic quantum dot. It is well-known that the charge pumped through chaotic quantum dots (and hence the output power of the corresponding quantum motor) vanishes with increasing phase breaking. (ii) Quantum motors can have ideal efficiency $\eta=1$, implying perfect conversion of electrical into mechanical energy. Indeed, this can be realized by motors based on Thouless pumps; when the Fermi energy lies in the gap, the conductance vanishes while the pumped charge is quantized to integer multiples of $e$. Thus, Eq.\ (\ref{Efficiency}) yields $\eta=1$, making Thouless pumps {\em ideal} adiabatic quantum motors. 

{\it Motor dynamics}.---The output power of a quantum motor depends on its dynamics through the cycle period $\tau$. Here, we discuss this for the simplest case, in which both the driving force and the load $F_{\rm load}$ acting on the angular motor degree of freedom $\theta$ are independent of the state of the motor. (This is realized for Thouless motors, but typically not for motors based on chaotic quantum dots.) If the motor degree of freedom is subject to damping with damping coefficient $\gamma$, the steady-state velocity of the motor follows from the (classical) condition
\begin{equation}
  \gamma\dot \theta = \frac{Q_p V}{2\pi} - F_{\rm load}.
  \label{EOM}
\end{equation}
Thus, we obtain for the cycle period of the motor $\tau = {2\pi}/{|\dot \theta|} = {(2\pi)^2\gamma }/({Q_pV - 2\pi F_{\rm load}})$.
We can use Eq.\ (\ref{current}) to eliminate $V$ in favor of the current $\overline{I}$. This yields
\begin{equation}
  \frac{1}{\tau} = \frac{Q_p\overline{I} - 2\pi F_{\rm load} \overline{G}}{Q_p^2 + (2\pi)^2 \gamma  \overline{G}}.
  \label{period}
\end{equation}
For an ideal Thouless motor with $\overline{G}=0$ \cite{Qi09}, this yields the relation $1/\tau = \overline{I}/Q_p$. This is a direct consequence of the fact that in this case, the entire current passing the device must be due to pumping. More generally, this remains a good approximation as long as $\overline{G} \ll Q_p^2/(2\pi)^2\gamma $. This result also implies that the maximum load on the motor is given by $F_{\rm load}^{\rm max} = Q_p\overline{I}/2\pi\overline{G}$.  

{\em Thouless motor.---}Thouless motors provide an instructive example not only because they realize ideal quantum motors but also because they allow for a thorough analytical discussion. Consider a single-channel quantum wire subject to a periodic potential of period $a$, as described by the Hamiltonian 
\be
 \H = {p}^2/2m + 2\Delta \cos (2\pi x/a + \theta) \Theta(L/2-|x|)
\label{TH}
\ee
The periodic potential of strength $2\Delta$ acting for $-L/2 < x < L/2$ arises, e.g., from a periodic set of charges situated along a conveyor belt or cogwheel so that the nearby electrons in the wire experience an electrostatic potential [cf.\ Fig.\ \ref{genericmotor}(b)]. This potential slides as the cogwheel turns and the mechanical variable $\theta$ varies by $2\pi$ as the teeth of the cogwheel advance by one spacing $a$. 

When the chemical potential $\mu$ is chosen such that the Fermi wavevector $k_F=({2m\mu/\hbar^2})^{1/2}$ is close to $k_0=\pi/a$, one can linearize the Hamiltonian for momenta close to $\pm k_0$. This results in an effective Hamiltonian $\H$ with counterpropagating linear channels and backscattering due to the periodic potential. Measuring momenta from $\pm k_0$ and energies from $\hbar^2 k_0^2/2m$, one has
\be
  \H = v_F  p \sigma^z + \Delta\l(\sigma^x \cos\theta + \sigma^y \sin\theta \rr) \Theta(L/2-|x|).
\label{TH2}
\ee
Here, the $\sigma^i$ denote the Pauli matrices in the space of the counterpropagating channels. We do not include the real electron spin for simplicity.

\begin{figure}[b]
\includegraphics[width=10.cm]{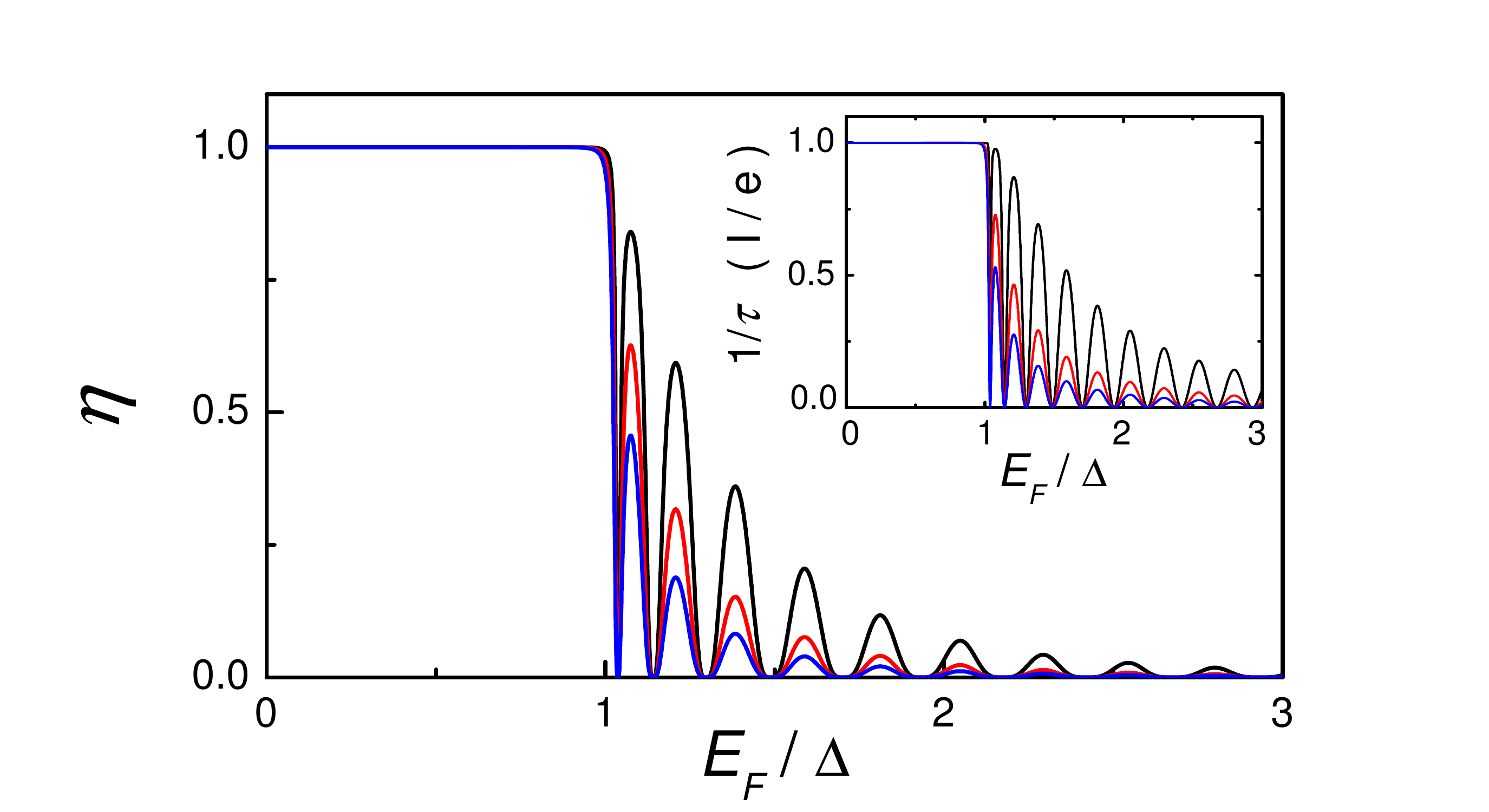}
\caption{Efficiency of the Thouless motor vs Fermi energy for $L=0.75\mu$m and $v=10^5$m/s. From top to bottom, the curves correspond to dissipative loads $\gamma/\hbar=1/2\pi,\,0.5,\,1$. The motor has ideal efficiency ($\eta = 1$) when the Fermi energy lies in the gap, $|E_F|<\Delta$, and the length is taken to infinity. The inset shows the cycle frequency for a current-biased Thouless motor vs Fermi energy. 
\label{efficiency}}
\end{figure}

Within the linearized model, the adiabatic $S$-matrix $S(\theta)$ can be readily obtained analytically. We start with the transfer matrix $M$ from $x=L/2$ to $x=-L/2$. Since the model is linear in momentum $p$, this can be done by analogy with the time-evolution operator in quantum mechanics
which yields
\be
 M =\exp\left\{ -\frac{iL}{\hbar v_F}\sigma^z \left[E-\Delta (\sigma^x\cos\theta+\sigma^y\sin\theta)\right]\right\}.
\ee
This can be rewritten as ${M}=\cos\lambda_L-i\sigma_{\rm eff}\sin\lambda_L$, where $\sigma_{\rm eff} = [{E \sigma^z -i\Delta\cos\theta\sigma^y+i\Delta\sin\theta\sigma^x}]/[{E^2-\Delta^2}]^{1/2}$ and $\lambda_L=(L/\hbar v_F) [E^2-\Delta^2]^{1/2}$. Note that $\sigma_{\rm eff}^2={\bf 1}$.

To obtain the $S$-matrix from the transfer matrix $M$, we first assume that there is only an outgoing wave on the right. Then, the wavefunction on the left is $(i_L,o_L)^T =M (o_R, i_R)^T=\l(M_{11} o_R,M_{21} o_R\rr)^T$, where $i$ and $o$ refer to the in- and outgoing waves, respectively. This immediately implies that the transmission $S_{21}$ is $1/M_{11}$, and the reflection $S_{11}$ is $M_{21}/M_{11}$. Repeating the same arguments with only an outgoing wave on the left, we also find $S_{22}= (M^{-1})_{12}/(M^{-1})_{22}$ and $S_{12} = 1/(M^{-1})_{22}$. With $M_{11}=(M^{-1})_{22}=\cos\lambda_L-i\frac{E}{\sqrt{E^2-\Delta^2}}\sin\lambda_L$, this yields
\be
 S =\frac{1}{M_{11}}\l(\ba{cc} -i e^{i\theta} \frac{\Delta}{\sqrt{E^2-\Delta^2}}\sin\lambda_L & 1 \\
     1 & -i e^{-i\theta} \frac{\Delta}{\sqrt{E^2-\Delta^2}}\sin\lambda_L \ea \rr).
\ee
We can now use this $S$-matrix to obtain explicit expressions for the efficiency of the Thouless motor. 

Using the Landauer formula, the conductance for a Fermi energy $E_F$ takes the form 
\be
G = \overline{G}=\frac{e^2}{h}\frac{|\Delta^2-E_F^2|}{|\Delta^2-E_F^2|\cos^2\lambda_L+E_F^2|\sin\lambda_L|^2}
\label{conductance}
\ee
In accord with the fact that the periodic potential opens a gap, the conductance is exponentially small in $L$  for $|E_F|<\Delta$ and becomes oscillatory and finite in $L$ for $|E_F| > \Delta$. Similarly, we can obtain the pumped charge in the standard way from Brouwer's formula \cite{Brouwer} (evaluated at zero temperature and for an angular degree of freedom),
\be
 Q_p =  \frac{e\Delta^2 |\sin \lambda_L|^2}{|\Delta^2-E_F^2|\cos^2\lambda_L+E_F^2 |\sin\lambda_L|^2} .
\label{ThoulessCharge}
\ee
For Fermi energies in the gap, the charge pumped is quantized to $e$ with exponential precision. When the Fermi energy is outside the gap, $|E_F|>\Delta$, the charge is no longer quantized and smaller than $e$. 

We can combine these results to obtain an explicit expression for the efficiency of the Thouless motor. To do so, we note that the force acting on the motor is independent of $\theta$. Thus, we can combine Eqs.\ (\ref{Efficiency}), (\ref{period}), (\ref{conductance}), and (\ref{ThoulessCharge}) to obtain (for zero load, $F_{\rm load}=0$)
\be
\eta = \frac{1}{1 + \frac{2\pi \gamma}{\hbar} \frac{|E_F^2-\Delta^2|}{\Delta^4 |\sin\lambda_L|^4}[|E_F^2-\Delta^2| \cos^2 \lambda_L +E_F^2 |\sin\lambda_L|^2]          }
\label{eta1}
\ee
In Fig.\ \ref{efficiency}, we plot the efficiency of the Thouless motor as a function of the Fermi energy. As can be seen from Eq.\ (\ref{eta1}), the efficiency is exponentially close to unity when the Fermi energy is within the gap. For this range of Fermi energies, the Thouless motor is an ideal adiabatic quantum motor. When the Fermi energy moves out of the energy gap, the efficiency is oscillatory with an algebraically dropping amplitude. In this regime, Fabry-Perot interference alone produces peaks in the efficiency, which appear when the reflection coefficients are maximal. The inset of Fig.\ \ref{efficiency} also shows the cycle frequency of the Thouless motor, for a given current and zero load, as a function of Fermi energy, cf.\ Eq.\ (\ref{period}).

{\em Intrinsic damping.---}So far, we have treated the damping coefficient $\gamma$ of the motor degree of freedom as phenomenological. However, in addition to extrinsic, purely mechanical friction, there is a contribution to $\gamma$ which arises intrinsically from the coupling to the electronic system. As shown recently, this intrinsic damping $\gamma_{\rm int}$ can also be obtained from the electronic $S$-matrix \cite{Bode11,Bode12,Thomas12}. Restricting attention to small bias voltages, we can approximate $\gamma_{\rm int}$ by its equilibrium value, $\gamma_{\rm int} = (\hbar/4\pi){\rm tr}[(\partial S^\dagger/\partial \theta)(\partial S/\partial \theta)]$. This is readily evaluated for the Thouless motor when the Fermi energy is in the vicinity of the fundamental gap. We find that the intrinsic damping can be expressed in terms of the pumped charge, $\gamma_{\rm int} = (\hbar/2\pi e) Q_p$. Quite surprisingly, the electronic system induces finite mechanical damping even when the Fermi energy lies in the gap (and $Q_p=e$). We interpret this damping as arising from forming plasmon excitations in the leads when pumped charge enters or leaves. When the Fermi energy of a current-biased Thouless motor lies inside the fundamental gap, the motor (without load) rotates at angular frequency $\omega = 2\pi I/e$, which, from Eq.\ (\ref{EOM}), gives a friction-induced voltage drop of $V = I (2\pi)^2 \gamma /e^2 $. The existence of the intrinsic friction implies that for a given current, there is a minimal voltage of $V = (h/e^2)I$ at which the Thouless motor described by Eq.\ (\ref{TH2}) can operate.  

At first sight, the intrinsic damping may seem to negate the possibility of an ideal quantum motor when the motor is subject to a load. Indeed, the electrical input power is then split between the power consumed by the load, $P_{\rm load} = F_{\rm load}\dot\theta$, and the power dissipated by damping, $P_{\gamma} = \gamma {\dot \theta}^2$. Nevertheless, for a quantized Thouless pump, $\dot\theta = 2\pi I/Q_p$, so that $P_{\rm load} \propto I$ while $P_{\gamma} \propto I^2$. Hence, the power dissipated by damping becomes negligible at small currents, and the load efficiency $\eta_{\rm load} = P_{\rm load}/P_{\rm in}$ can be made arbitrarily close to unity by operating the motor at low currents.  

{\em Conclusions.---}Motion at the nanoscale tends to be dominated by fluctuations. It is an important challenge to develop schemes to generate directed motion in nanoscale devices \cite{Kudernac,Tierny,Perera}. Here, we investigated a general strategy to this effect which is based on operating quantum pumps in reverse. We developed a corresponding theory which expresses the output power and the efficiency of such adiabatic quantum motors to characteristics of the pumps on which they are based. The concept of adiabatic quantum motors offers numerous possibilities for future research. Interesting directions include motors based on electron pumps which involve electron-electron interactions as well as systems in which the motor degree of freedom is itself quantum mechanical.

We acknowledge discussions with P.\ Brouwer as well as support by the Deutsche Forschungsgemeinschaft through SFB 658, the Humboldt Foundation through a Bessel Award, the Packard Foundation, and the Institute for Quantum Information and Matter, an NSF Physics Frontiers Center with support of the Gordon and Betty Moore Foundation.


\begin{thebibliography}{12}

\bibitem{Feynman} R.P.\ Feynman, Engineering and Science {\bf 23}, 22 (1960).

\bibitem{Zettl} A.M.\ Fennimore, T.D.\ Yuzvinsky, W.-Q.\ Han, M.S.\ Fuhrer, J.\ Cumings, and A.\ Zettl, Nature {\bf 424}, 408 (2003).

\bibitem{Haenggi} A.V.\ Ponomarev, S.\ Denisov, and P.\ H\"anggi, Phys.\ Rev.\ Lett.\ {\bf 102}, 230601 (2009).

\bibitem{Weitz} T.\ Salger, S.\ Kling, T.\ Hecking, C.\ Geckeler, L.\ Morales-Molina, and M.\ Weitz, Science {\bf 326}, 1241 (2009).

\bibitem{Bailey08} S.W.D.\ Bailey, I.\ Amanatidis, and C.J.\ Lambert, Phys.\ Rev.\ Lett.\ {\bf 100}, 256802 (2008).

\bibitem{Dundas09} D.\ Dundas, E.J.\ McEniry, and T.N.\ Todorov, Nature Nanotech.\ {\bf 4}, 99 (2009).

\bibitem{Qi09} X.-L.\ Qi and S.C.\ Zhang, Phys.\ Rev.\ B {\bf 79}, 235442 (2009).

\bibitem{Brouwer} P.W.\ Brouwer, Phys.\ Rev.\ B {\bf 58}, R10135 (1998).

\bibitem{Switkes} M.\ Switkes, C.M.\ Marcus, K.\ Campman, A.C.\ Gossard, Science {\bf 283}, 1905 (1999).

\bibitem{Lu10} J.T.\ L\"u, M.\ Brandbyge, and P.\ Hedegard, Nano Lett.\ {\bf 10}, 1657 (2010). 

\bibitem{Bode11} N.\ Bode, S.\ Viola Kusminskiy, R.\ Egger, F.\ von Oppen, Phys.\ Rev.\ Lett.\ {\bf 107}, 036804 (2011).

\bibitem{Bode12} N.\ Bode, S.\ Viola Kusminskiy, R.\ Egger, F.\ von Oppen, Beilstein J.\ Nanotechnol. {\bf 3}, 144 (2012).

\bibitem{Thomas12} M.\ Thomas, T.\ Karzig, S.\ Viola Kusminskiy, G.\ Zarand, F.\ von Oppen, Phys. Rev. B {\bf 86}, 195419 (2012).

\bibitem{Berry} M.\ V.\ Berry, in {\em Geometric Phases in Physics}, edited by A. Shapere and F. Wilczek (World Scientific, Singapore, 1989).

\bibitem{Thouless} D.J.\ Thouless, Phys. Rev. B {\bf 27}, 6083 (1983).

\bibitem{Kudernac} T.\ Kudernac,	N.\ Ruangsupapichat,	M.\ Parschau, B.\ Macia, N.\ Katsonis, S.R.\ Harutyunyan,	K.-H.\ Ernst, and B.L.\ Feringa, Nature {\bf 479}, 208 (2011).

\bibitem{Tierny} H.L.\ Tierney, C.J.\ Murphy,	A.D.\ Jewell, A.E.\ Baber, E.V.\ Iski, H.Y.\ Khodaverdian, A.F.\ McGuire, N.\ Klebanov, and E.C.H.\ Sykes, Nature Nanotech.\ {\bf 6}, 625 (2011).

\bibitem{Perera} U.G.E.\ Perera, F.\ Ample, H.\ Kersell, Y.\ Zhang, G.\ Vives, J.\ Echeverria, M.\ Grisolia, G.\ Rapenne, C.\ Joachim, and S.-W.\ Hla
Nature Nanotech. {\bf 8}, 46 (2013).

\end{thebibliography}
\end{document}